\documentclass{elsart}
\usepackage{amssymb}
\usepackage{graphicx}
\begin{document}
\begin{frontmatter}
\title{\Large \bf Dynamic Anomalies
  of 
  Fluids with  Isotropic Doubled-Ranged Potential}
  \author[ulbra,lasale]{Paulo A. Netz}
  \ead{netz@iq.ufrgs.br}
  \author[ufrgs]{Jos\'e Fernando Raymundi}
  \ead{raymundi@if.ufrgs.br}
  \author[lasale]{Adriana Simane Camera}
  \ead{raymundi@if.ufrgs.br}
  \author[ufrgs]{Marcia C. Barbosa}
  \ead{barbosa@if.ufrgs.br}
  \ead[url]{ http://www.if.ufrgs.br/$\sim$barbosa}
  \address[ulbra]{Departamento de Qu\'{\i}mica, Universidade Luterana do
  Brasil \\ 92420-280, Canoas, RS, Brazil}
 \address[lasale]{ Canoas, RS, Brazil}
  \address[ufrgs]{Instituto de F\'{\i}sica, Universidade Federal do Rio
  Grande do Sul\\ Caixa Postal 15051, 91501-970, Porto Alegre,
  RS, Brazil}
  \begin{abstract}
Using molecular dynamics simulations we investigate
the ability of  an analytical three-dimensional double well
in reproducing static and dynamic anomalies found experimentally
in liquid water. We find anomalous behavior in 
the stable region of the phase
diagram   if the outer minimum is deeper than the inner minimum. In
the case of a deeper inner minimum, anomalous behavior 
is also present but inside the unstable region.

  \end{abstract}
  \end{frontmatter}

  \section{Introduction}
  \label{1}

  The prediction of structural, dynamic and thermodynamic properties
  and the phase behavior of fluids based solely on the shape of the 
  intermolecular interaction potential is an important goal in the
  study of liquids and in the understanding of the peculiarities of
  the liquid state. The simplicity of this surmise   has  not only 
  theoretical interest but also practical consequences\cite{jag98}.
  In spite of these motivations and precepts of developing a much
  better molecular-level picture of the liquid matter, this goal is
  far from being achieved.  Although for some simple potential the use of integral
  equations in the form of hypernetted-chain 
  approximation can give an  estimate of the phase
  stability \cite{Ca96}, the mathematical complexity of the interaction
  potentials used even for simple liquid  does not allow an analytical 
  approach.  Despite of this, some
  progress can  be made choosing only
  liquids that share the same  type of  structure, for instance, a 
tetrahedrical local structure. 
  The most familiar tetrahedral  liquid is water
  not only due to its abundance but also because
  of the occurrence of some remarkable
  features such as the presence of structural, dynamic
  or thermodynamic anomalies. The question is to determine
  which characteristics of the intermolecular interaction potential 
  would be needed to induce anomalies in the liquid phase behavior of 
  fluids governed by this potential. One of the most 
  familiar mysteries in water's behavior
   is the density anomaly: at 
   sufficient low temperatures,
   warming the liquid causes it to shrink.  
   There is an increasing awareness that including this one
   property in the potential will reveal the origins of the others.

   Recently, following
   an earlier hypothesis of Kamb \cite{Ka60}, and supported by 
experimental results \cite{Ro96}-\cite{Ch97}
   it has been shown that the increase in density
   on melting is related to a distortion in the
   non-nearest-neighbor coordination, i.e. bending of hydrogen bonds.
   The key point is therefore
   modeling the behavior of the system based not on a potential mimicking
   the first neighbors interactions, as in mostly current models, but on a 
   potential capable of describe the peculiar interactions between second
   neighbors \cite{cho96}\cite{cho97a}. These interactions are double-ranged:
the competition between two scales is responsible for many of 
the anomalies of water. 
  In this spirit one can understand
   the success of a simple model as  the core-softened  potential introduced
   by Stell et al. \cite{hem70}. It was 
   shown that an one-dimensional hard-core system showing a first
   order phase transition could be capable of exhibiting two (or even
   three) phase transitions if the hard-core is softened in a reasonable
   way. It is, however, not clear if the same essential features of this 
   model would produce anomalies in three-dimensional systems.

    Another example is the
     double-well Takahashi model\cite{cho96}, with 
     an inner shallow well and a outer  deeper well.  This model exhibits 
     a density maximum in one-dimensional systems, with the correct
     pressure dependence: this density maximum becomes broader and shifts to
     lower temperatures by increasing the pressure. This approach, however,
     seems not to work in three-dimensional systems\cite{cho97a,vel97} and 
     the question was posed if the three-dimensional potential, besides
     being double-ranged, also must have an angular dependence in order
     to show anomalies.  Cho and co-workers
     \cite{cho96b} also proposed an analytic double-well
     potential without angular dependence,
with a slight modification on a Lennard-Jones potential,
     adding a Gaussian-shaped perturbation.
     This potential seems to have anomaly in the density. However,  a more
     consistent exploration of the parameter space is necessary in order
     to determine if this is the case for all range of parameters.
     Later, Cho et al. \cite{cho97b} argued that only non local potentials 
     could reproduce all water's anomalies. 
Within their analysis the potential should  have a strong
     angular dependence and favor the build-up of a 
second-neighbor structure
\cite{Fr02}.

     Following a similar  approach as the original Takahashi model, 
     Jagla showed\cite{jag99} that several anomalies could
be described by a simple interaction potential with two competing
equilibrium distances. The model including a global term for attractions
     displays a liquid phase with a first
     order line of liquid-gas transition ending in a critical point.
     This potential, however, is difficult to implement in a molecular
     dynamics simulation, because the attraction term is not
     explicitly taken into account and the system needs long equilibration
     times. Besides the three models mentioned above, a number of lattice 
     \cite{De91}
     and continuous \cite{Sc01} core-softened  potentials 
     has been proposed.

      However, besides the thermodynamic,
      water also has dynamic anomalies.  While for most
      materials diffusivity decreases with pressure,
      liquid water has an opposite behavior in a large region
      of the phase diagram \cite{Pr87}-\cite{Ne01}.
      A good model for water and tetrahedral liquids should not only
      exhibit the thermodynamic but also the anomaly in the mobility.
These dynamical anomalies are related to the competition between 
the local tetrahedral structure of first neighbors - which tends to 
decrease the mobility and the distortion of the structure of first
and second neighbors, with weakened hydrogen bonds and interstitial 
water molecules - which tends to increase the mobility. 
This anomaly in the diffusivity 
is enhanced at low temperatures where the two scales phenomena
becomes more proeminent.
       In the present paper, we check if  a simple model in
which the particles are model as modified soft spheres interacting with  two
       competing ranges exhibits the anomalies present in
water. Particularly we study the diffusivity of 
 two  model cases: one in which the outer  minimum is
deeper and another in which the inner  minimum is deeper. 
      This paper is organized  as follows.
       In next section  the model and method are described and
       in sec. 3 the results and the conclusions are shown.

\begin{figure}
\begin{minipage}[t]{140mm}
\includegraphics[width=10cm,height=6cm]{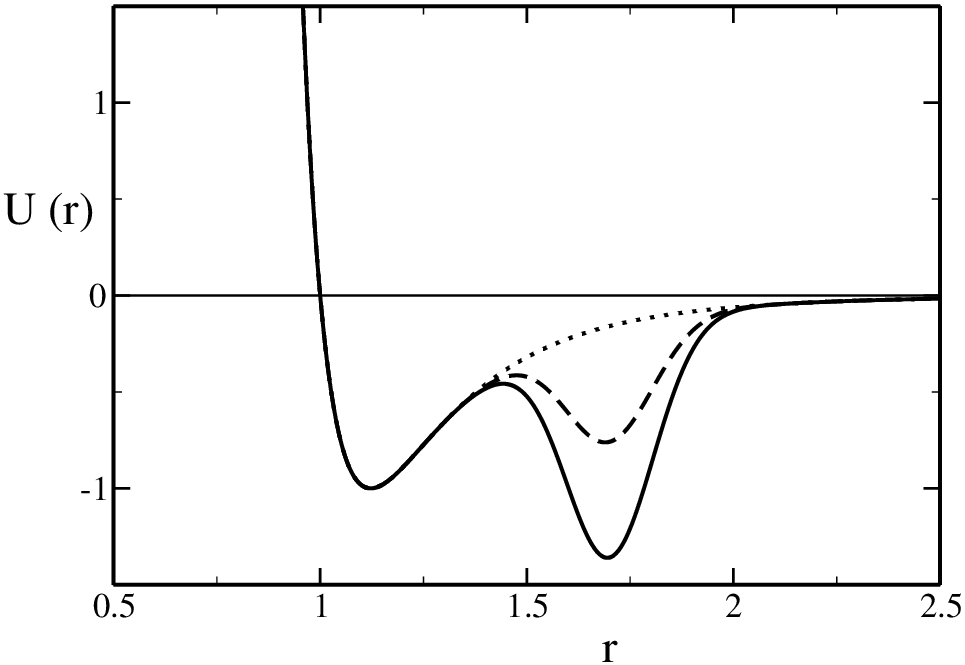}
\caption{Model potential I (solid line) with a outer deeper minimum and
parameters  $\alpha$ =- 1.2, $\beta$ = 1.7, $\gamma$ = 0.15, 
Model potential II (dashed line) with a inner deeper minimum,
and parameters $\alpha$ =- 0.6, $\beta$ = 1.7, $\gamma$ = 0.15, compared
with a normal Lennard-Jones potential (dotted line)}
\label{fig:fig1}
\end{minipage}
\end{figure}


       \section{The Model and the Methods}
       \label{2}

       Our model system consists of softened spheres with
an  intermolecular potential displaying 
       a repulsive short-ranged component showing 
       $r^{-12}$ dependence and a double-ranged attractive
       component, by adding a Gaussian to a normal Lennard-Jones 
       potential \cite{cho96b}, as follows:
\begin{equation}
U = 4 \left[ r^{-12} - r^{-6} \right] + \alpha \exp \left( - \frac{(r-\beta)^2}{\gamma^2} \right) 
\end{equation}
       This potential can represent a whole family of intermolecular
       interactions, from weak repulsive shoulder to deep double
       well potentials, depending on the choice of the values
       of $\alpha$, $\beta$ and $\gamma$. Figures 1-2
show two choices of
       parameters that illustrate two significative cases: one
in which the outer minimum is deeper and another in which the
inner minimum is minimum.
Depending on the choice of parameters, we can obtain  either a potential
similar to the one employed by Cho\cite{cho96}  
or the one introduced by Jagla\cite{jag99}.

We carried out molecular dynamics simulations in the microcanonical
ensemble of 256 particules interacting with an
intermolecular potential given by the expression above.
A broad range of thermodynamic conditions, expressed as several
densities  and temperatures was  chosen to explore the
phase diagram of the fluid described by this potential.
Using reduced units for density $\rho^{*} = \rho\sigma^3$,
for temperature $T^{*} = T k_B \epsilon^{-1}$, the values for
$\rho^{*}$ and $T^{*}$ were 0.60 to 1.00 and 0.30 to 1.00 respectivelly.
Thermodynamic parameters (internal energy and pressure) were calculated
over 3000000 steps long simulations. Analysis of the dependence of
the pressure against density along isotherms, aided by visual inspection 
of the final structure allows to check if the system is unstable or not.
We measure the mean-square displacement averaged over different initial
times
$\langle \Delta r^2(t) \rangle = \langle [r(t+t_0)-r(t)]^2\rangle$
       and then the diffusion constant is calculated  using the relation
       $D=\langle \Delta r^2(t) \rangle/6t$.

       \section{Results and Conclusions}
       \label{3}

       The diffusion constant  is
       calculated for   two types of potentials:
  (I)$\alpha = -1.2$, $\beta = 1.7$ and 
       $\gamma = 0.15$, (II)  $\alpha = -0.6$, $\beta = 1.7$ and
       $\gamma = 0.15$.
        Potential I   exhibit  a competition between a first attractive
       shell and a second deeper  minimum. 
        This competing interactions  can generate 
	two liquid phases with different densities \cite{cho96b}\cite{De91}\cite{Sc01}
	and a resulting density anomaly.  
 The mobility of potential I, illustrated in 
	Figure 2,  is similar to the behavior  
	we found \cite{Ne01} in SPCE \cite{Be87} supercooled water.
	The diffusivity increases as the density is lowered, reaches
	a maximum and decreases. It is not clear if for this potential
the diffusivity increases at very low densities \cite{Ne01} because
the system cavitates before this happens. 
	 Potential II has a competition between a deeper inner well and
	 an shallow outer shell. This competition can also generate two liquid
	 phases with different densities \cite{De91} and 
	 a resulting density anomaly. The 
	 diffusion constant related to Potential II, illustrated in
	 Figure 3, is also anomalous, however its anomalous behavior
occurs inside the unstable region where the system already is phase
separated.

	 According Cho et. al. \cite{cho96}\cite{Sc01} a double well Takahashi
	 model would exhibit a density maximum only if the outer well
	 is deeper than the inner well.
	 Although our potential is not of a Takahashi type, its shape
	 roughly resembles that potential by an adequate choice of 
parameters.  Our main conclusion is that even if   the 
outer well is shallower than the inner well   the anomalies are still
present, however they are covered by the phase separation. A systematic study in progress of the dependence of the fluid
	 behavior  with the  choice of parameters will  enable us 
	 to better  understand  the hierarchy of the
	 several anomalies in such a two-ranged analytic model.

\begin{figure}
\begin{minipage}[t]{70mm}
\includegraphics[width=6cm,height=5cm]{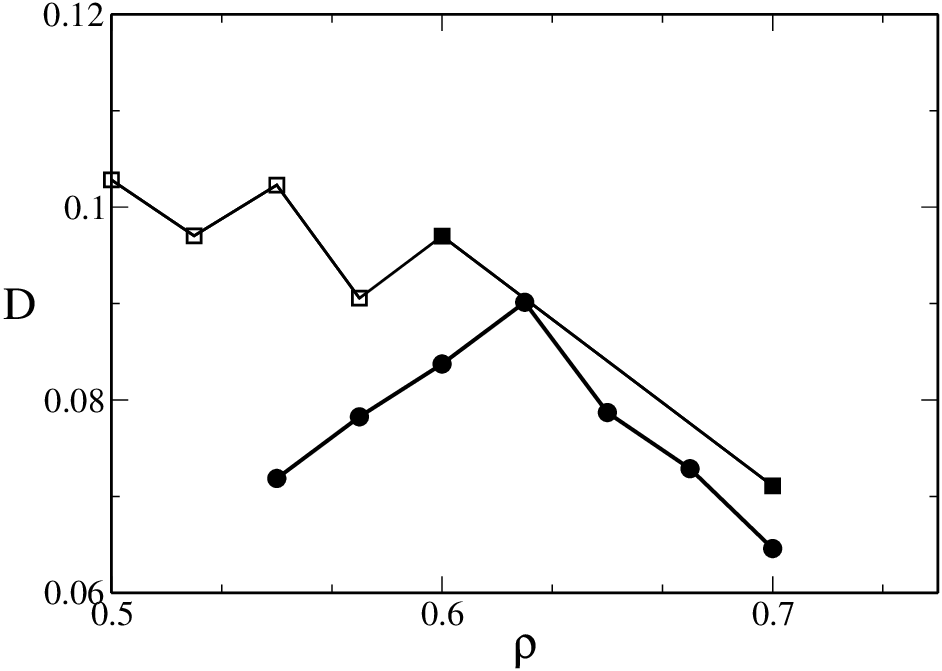}
\caption{Diffusion coefficient as a function of density for 
temperatures T = 0.70 (circles) and T = 0,75 (squares)
using Potential I. All
quantities are in reduced units. Open symbols correspond to 
systems where there is cavitation.}
\label{fig:fig3}
\end{minipage}
\hspace*{0.5cm}
\begin{minipage}[t]{70mm}
\includegraphics[width=6cm,height=5cm]{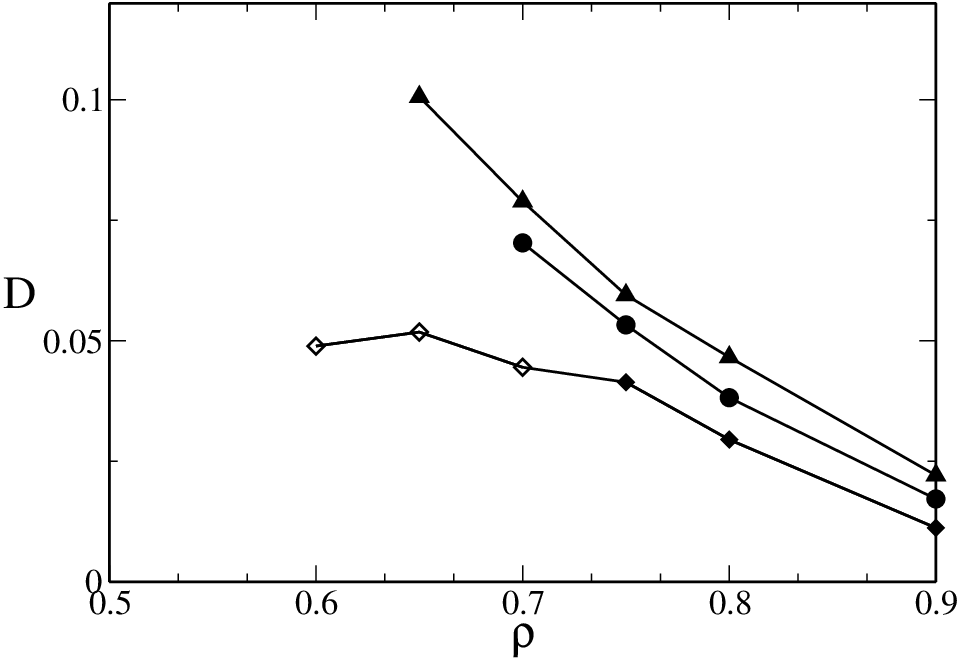}
\caption{Diffusion coefficient as a function of density for 
temperatures T = 0.60 (diamonds), T = 0.70 (circles) and  T = 0.80 (triangles),
using Potential II, same conventions as the figure before.}
\label{fig:fig4}
\end{minipage}
\end{figure}

		  \end{document}